\pdfoutput=1

\documentclass[11pt]{article}

\usepackage[final]{acl}

\usepackage{times}
\usepackage{latexsym}
\usepackage{fontawesome}
\usepackage{amsmath}

\usepackage[T1]{fontenc}

\usepackage[utf8]{inputenc}

\usepackage{microtype}

\usepackage{inconsolata}

\usepackage{graphicx}

\usepackage[breakable]{tcolorbox}
\usepackage{algorithmic}
\usepackage{algorithm}
\usepackage{booktabs}
\usepackage{multirow}

\title{The Stepwise Deception: Simulating the Evolution from \\ True News to Fake News with LLM Agents}

\author{
   Yuhan Liu\textsuperscript{1}, Zirui Song\textsuperscript{2}, Juntian Zhang\textsuperscript{1}, Xiaoqing Zhang\textsuperscript{1}, Xiuying Chen\textsuperscript{2}\thanks{\ \ Corresponding authors.},
  Rui Yan\textsuperscript{1,3,4}\footnotemark[1] \\
  \textsuperscript{1}Gaoling School of Artificial Intelligence, Renmin University of China,  \textsuperscript{2}MBZUAI, \\ \textsuperscript{3}Engineering Research Center of Next-Generation Intelligent Search and Recommendation, MoE\\
  \textsuperscript{4}School of Artifcial Intelligence, Wuhan University\\
  \texttt{yuhan.liu@ruc.edu.cn}
}

\begin{document}
\maketitle
\begin{abstract}
With the growing spread of misinformation online, understanding how true news evolves into fake news has become crucial for early detection and prevention. However, previous research has often assumed fake news inherently exists rather than exploring its gradual formation. To address this gap, we propose \textbf{FUSE} (\textbf{F}ake news evol\textbf{U}tion \textbf{S}imulation fram\textbf{E}work), a novel Large Language Model (LLM)-based simulation approach explicitly focusing on fake news evolution from real news. Our framework model a social network with four distinct types of LLM agents commonly observed in daily interactions: \textit{spreaders} who propagate information, \textit{commentators} who provide interpretations, \textit{verifiers} who fact-check, and \textit{bystanders} who observe passively to simulate realistic daily interactions that progressively distort true news. To quantify these gradual distortions, we develop \textbf{FUSE-EVAL}, a comprehensive evaluation framework measuring truth deviation along multiple linguistic and semantic dimensions. Results show that FUSE effectively captures fake news evolution patterns and accurately reproduces known fake news, aligning closely with human evaluations. Experiments demonstrate that FUSE accurately reproduces known fake news evolution scenarios, aligns closely with human judgment, and highlights the importance of timely intervention at early stages. Our framework is extensible, enabling future research on broader scenarios of fake news.
\end{abstract}

\section{Introduction}
\begin{figure*}[tb]
    \centering
    \includegraphics[width=0.9\linewidth]{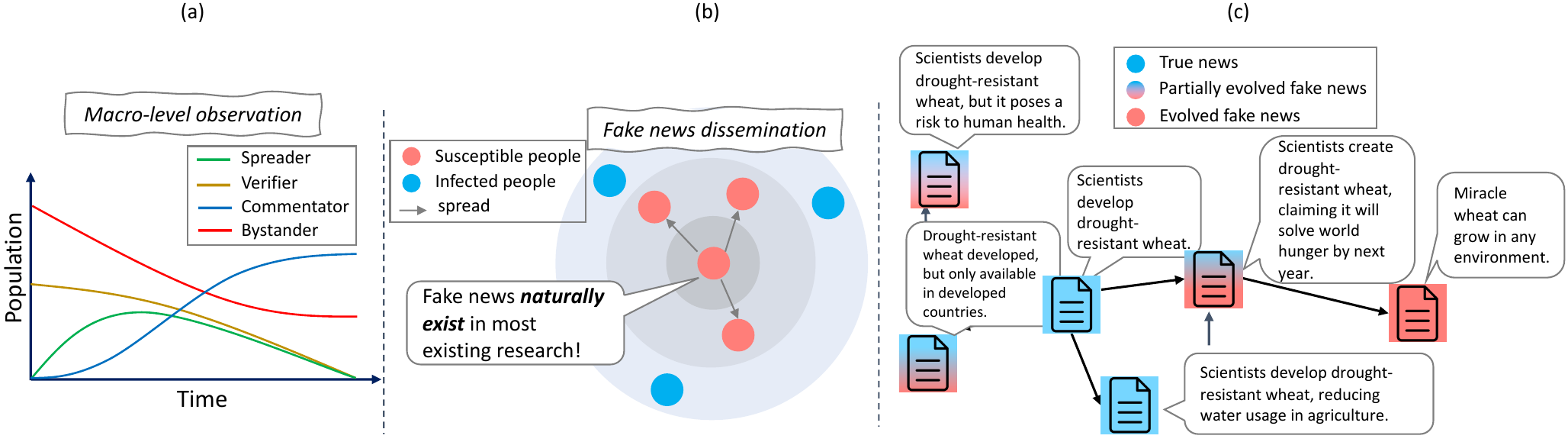}
    \caption{
    (a) Macro-level observation of population dynamics based on the mathematical model, categorizing individuals into four types and showing their quantity changes over time. 
    (b) The micro-level conventional fake news dissemination model assumes that fake news inherently exists.
    (c) Micro-level evolution of fake news, where true news gradually evolves into fake news during network propagation with content alterations at various stages.}
    \label{fig:intro}
\end{figure*}

The rapid spread of fake news has become a significant global concern~\cite{Lazer2018TheSO,Olan2022FakeNO}.
Prior research predominantly addresses fake news detection or simulates the spread of misinformation after its initial appearance~\cite{garimella2017balancing,wang2019efficient}. 
For instance, \citet{Piqueira2020DaleyKendalMI} categorized individuals into four types and used mathematical models to simulate the spread of fake news, as depicted in Figure~\ref{fig:intro}(a). 
On a micro-level, \citet{jalili2017information} defined numerical conditions for opinion change to study fake news dissemination, as shown in Figure~\ref{fig:intro}(b).
However, these models typically assume fake news as inherently existing entities within social networks, ignoring how misinformation originates or evolves over time.

In contrast, fake news may originate from true news that becomes distorted or misinterpreted over time, eventually evolving into fake news~\cite{guo2021does,shen2024online} as illustrated in Figure~\ref{fig:intro}(c). This evolutionary process is critically underexplored despite its significance for effective early interventions. Recognizing this gap, our work explicitly adopts the definition of fake news from prior influential research~\cite {lazer2018science}, focusing specifically on scenarios where factual information incrementally transforms into misinformation during its dissemination (Figure~\ref{fig:intro}(c)). We define this transitional content as \textit{partially evolved fake news}, characterized by a mix of accurate and distorted elements.

Specifically, we propose the Fake news evolUtion Simulation framEwork (\textbf{FUSE}), the first comprehensive approach employing LLM agents to simulate how real news progressively evolves into fake news within different social network structures (e.g., high-clustering, scale-free, and random networks). The simulation consists of four distinct agent roles commonly found in real-world interactions: \textit{spreaders}, who disseminate information; \textit{commentators}, who interpret content; \textit{verifiers}, who assess factual accuracy; and \textit{bystanders}, who observe without active participation. Each agent engages daily, exchanging beliefs, reevaluating information, and contributing to incremental content distortions. Our agents incorporate hierarchical memory structures, combining short-term interactions and long-term knowledge, allowing realistic reflective reasoning processes and dynamic content adaptation.

Given the absence of prior work on language-based evaluation of fake news evolution, we introduce \textbf{FUSE-EVAL},  
a novel multidimensional evaluation framework that quantifies the deviation of evolved news from its original form across multiple dimensions, including Sentiment Shift (SS), New Information Introduced (NII), Certainty Shift (CS), STylistic Shift (STS), Temporal Shift(TS), and Perspective Deviation (PD).
Our comprehensive experiments validate FUSE's \textit{strong alignment} with real-world observations from prior research.
The results reveal three key findings: 
(1) news exhibits \textit{clear accumulation distortion effects}, where content progressively deviates from its original form during spread~\cite{de2024application};
(2) true news evolution to fake news occurs more rapidly in \textit{high-clustering networks} than in scale-free or random networks~\cite{trpevski2010model};
(3) \textit{political news} shows significantly faster evolution rates compared to other topics (terrorism, natural disasters, science, and finance)~\cite{lazer2018science}.

To construct a responsible online environment, our research reveals the importance of \textit{strategic interventions during the early stages of fake news evolution}.
Rather than waiting until fake news has widely spread, 
we introduce an official agent that intervenes when information deviation reaches critical thresholds, issuing authoritative statements with reliable sources to counteract misinformation spread.
This early intervention approach demonstrates the effectiveness of timely, authoritative responses in misinformation governance.

Our contributions can be summarized as follows:

 $\bullet$ \textit{\textbf{Versatile Framework.}} We propose FUSE, an LLM-based simulation framework to investigate how true news gradually evolves into fake news, and validate through experiments that our framework successfully reproduces real-world phenomena by considering different types of agents and various social network structures.

 $\bullet$ \textit{\textbf{Comprehensive Evaluation.}} We introduce FUSE-EVAL, a novel multidimensional framework to measure the deviation from true news during news evolution.

 $\bullet$ \textit{\textbf{Practical Insights.}} We propose and evaluate multiple intervention strategies aimed at mitigating the spread of fake news during its evolution.

\section{Related Work}

\paragraph{Fake News Evolution}
Recent research into fake news evolution has focused on how misinformation spreads and transforms over time. 
\citet{zhang2013rumor} found that rumors evolve as they are repeatedly modified, becoming shorter and more shareable, while \citet{guo2021does} empirically tracked fake news evolution, noting how sentiment and text similarity change as truth transitions into misinformation. \citet{xia2020state} proposed a sentiment analysis pipeline to track public opinion shifts in fake news by detecting sarcasm.
Other studies have emphasized structural and behavioral aspects of fake news propagation. 
\citet{zhao2024propagation} proposed a dynamic method that captures temporal changes in rumor propagation, revealing how rumor patterns evolve. 
\citet{wang2021evolution} demonstrated slight news content changes during the COVID-19 pandemic, while \citet{li2016user} examined how user behaviors, particularly the role of verified accounts, influence the evolution of rumors. FPS~\cite{liu2024skepticism} and TED~\cite{liu2025truth} uses a multi-agent system to study the propagation and detection of fake news.

However, there has not been a detailed and comprehensive study on how true news evolves into fake news, with only some superficial linguistic analyses~\cite{zhang2013rumor,guo2021does}.

\paragraph{LLMs as Agents}

Agent-based modeling simulates complex systems through individual agents' interactions in dynamic environments~\cite{macal2005tutorial}. The integration of LLMs has enhanced these simulations by enabling natural language processing capabilities~\cite{zhang2025weaving,chen2023improving,chen2023topic} and human-like intelligence in planning and decision-making~\cite{xi2023rise}. This has led to widespread adoption across various domains~\cite{li2023you,park2023generative,liu2025truth,jin2025beyond}, establishing LLM agents as a new paradigm for human-level intelligence simulation.
In more specific applications, LLM agents have been employed to simulate social media dynamics. 
For instance, \citet{tornberg2023simulating} used them to investigate social media algorithms and provide insights into real-world phenomena, while \citet{park2022social} demonstrated their ability to generate human-like social media content. 
Our work extends this approach by being one of the first to apply LLM agents in simulating fake news evolution. 

\section{Methodology}
\label{sec:method}
\begin{figure*}[ht]
    \centering
    \includegraphics[width=0.8\linewidth]{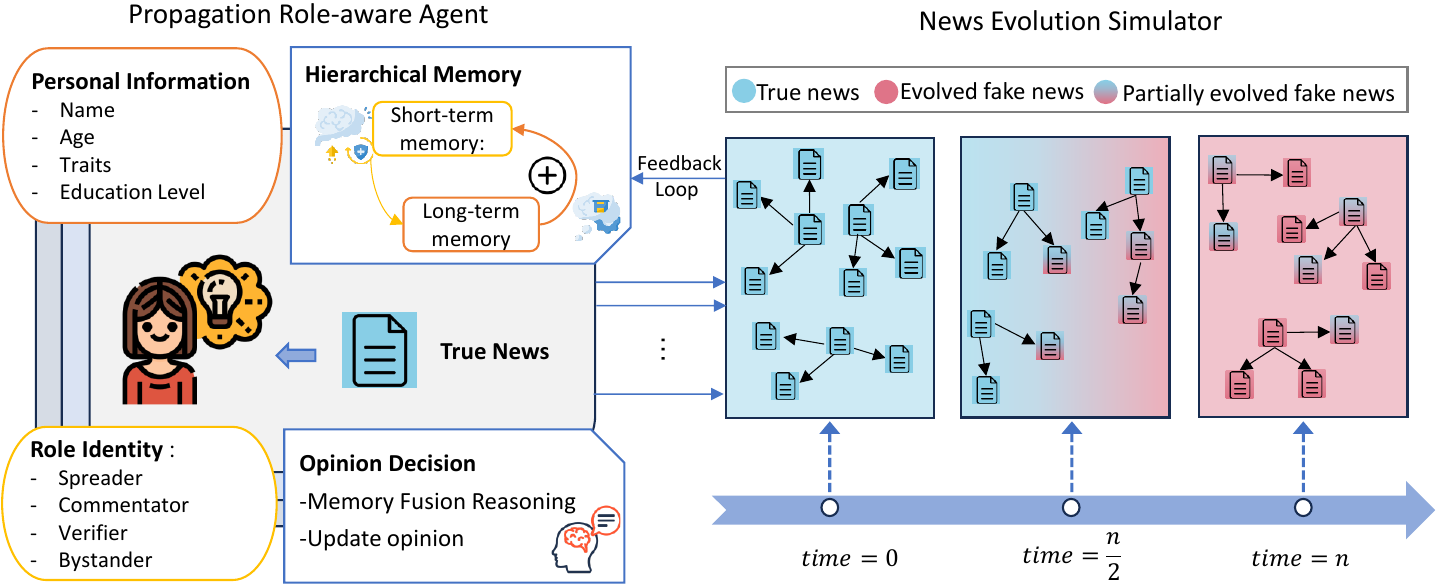}
    \caption{Our FUSE framework simulates news evolution by equipping each agent with role-based decision-making capabilities. Propagation Role-aware agents (PRA) process true news through interactions within the news evolution simulator (NES), where their role identities shape how they engage with the news.
    }
    \label{fig:model}
\end{figure*}

\subsection{Problem Formulation}
We simulate the gradual evolution of true news into fake news using LLMs as agents within a social network, consistent with the definition of fake news provided by prior research~\cite{lazer2018science, guo2021does}.  
The simulation consists of $N$ agents $\mathcal{A} =( a_1, \ldots, a_N)$, each endowed with a unique persona defining their Behavior role, personality traits, and demographic information.

At time $t=0$, true news $S_0$ is introduced into the network. 
The agents are connected according to a predefined social network structure  $\mathcal{G} = (\mathcal{A}, \mathcal{E})$, which may represent high-clustering, scale-free, or random networks to reflect real-world dynamics.
On each day $t = 1, 2, \ldots, T$, agents interact with their neighbors $\mathcal{N}_i$, exchanging information and opinions based on their personas and prior knowledge. 
After interactions, agents process and reintroduce the news content based on their updated beliefs.
The evolution of the news content for agent $a_i$ with a personal profile $\mathcal{P}_i$ at time $t$, denoted as $S_i^{t}$, is defined by:
\begin{equation}
    S_i^{t}=f(S_i^{t-1}, \{S_j^{t-1}|a_j \in \mathcal{N}_i \},\mathcal{P}_i),
    \label{eq:1}
\end{equation}
where $f(\cdot)$ represents the agent's information processing function.

Through this simulation, we analyze how the true news $S_0$ transforms over time due to agents' interactions and personal biases, examining the impact of agent types, network structures, and individual traits on the evolution of fake news.

\subsection{Our Simulation Framework}
As depicted in Figure~\ref{fig:model}, our FUSE framework consists of two core components: the \textit{Propagation Role-Aware agents} (PRA) and the \textit{News Evolution Simulator} (NES). 
The PRA module empowers agents with role-based decision-making capabilities, while the NES establishes the interaction environment, simulating the social network through which news propagates and evolves.
Within the PRA module, each agent is powered by an LLM and characterized by a specific role type and personal attributes, which govern their information processing, interaction patterns, and opinion updates.  
The NES facilitates daily interactions through a predefined social network structure, $\mathcal{G} = (\mathcal{A}, \mathcal{E})$,  simulating various network types to reflect different social dynamics. 

During each simulation day, agents engage with their network neighbors, exchanging news content and opinions shaped by their roles and attributes. When news content deviates beyond a set threshold, intervention mechanisms—such as official announcements—are triggered to provide credible information and correct potential misinformation. The simulation advances daily with updated agent states, tracking the evolution of news content through the network.

\subsection{Propagation Role-Aware Agent} 
\label{lab:PRA}

The PRA is designed to simulate individual human behaviors in news evolution by equipping agents with specific roles and personal attributes, aiming to mirror the diversity and complexity of human interactions in social networks.

\subsubsection{Personal Information}
According to \citet{sun2023fighting}, the roles in fake news propagation can be classified into four types: \textit{spreaders}, who propagate information; \textit{commentators}, who provide opinions and interpretations; \textit{verifiers}, who check the accuracy of information; and \textit{bystanders}, who passively observe without engaging. 
However, they failed to model this in their numerical simulation.
We follow this setup but enhance it by equipping each agent \( a_i \) with a textual role description \( r_i \in \{ \textit{\text{spreader, commentator, verifier, bystander}} \} \).
Additionally, agents possess a personal profile $\mathcal{P}_i$ that includes demographic attributes (name, age, gender, and education level) and personal traits based on the Big Five model~\cite{barrick1991big}, which influence their information processing behaviors.

\subsubsection{Role-Specific Behaviors}

At each time step $t_i$, agent $a_i$ holds a version of the news content $S_i^t$. When interacting with neighboring agents $\mathcal{N}_i$ as defined by the network $\mathcal{G}$, agent $a_i$ receives news content $\{S_j^{t-1}|a_j \in \mathcal{N}_i \}$.
The agent then reintroduces news based on their role and persona through a role-specific update function:
\begin{equation}
    f_{role} = f_{r_i}(S_i^{t-1}, \{S_j^{t-1}|a_j \in \mathcal{N}_i \},\mathcal{P}_i).
\end{equation}
For different roles in our model, \textit{spreaders} may combine and amplify sensational aspects of the news, \textit{commentators} may add personal opinions, \textit{verifiers} may check news before sharing, and \textit{bystanders} may retain their previous news content unless significantly influenced~\cite{sun2023fighting}.

\subsubsection{Memory and Reflection}

 In our simulation, agents engage with their neighbors daily, leading to updated versions of the news. Given the volume of interactions, we implement a hierarchical memory system comprising short-term memory (STM) $M_i^{S}$ for recent interactions and long-term memory (LTM) $M_i^{L}$ for accumulated knowledge.  
 After interactions, agents reflect and update the news through a memory function:
\begin{equation}
\label{eq:3}
    M_i^{L,t} = g(f_{L}(M_i^{L,t-1}),f_{S}(M_i^{S,t})),
\end{equation}
where $g(\cdot)$ integrates new information into LTM, enabling agents to exhibit dynamic behaviors such as gradually changing their opinion on a topic or reinforcing existing opinions.

\subsubsection{Decision-Making Process}

In our FUSE framework, each agent's opinion evolves through a reasoning process influenced by their role, persona, and interactions. Agents reflect on their news content after daily interactions and memory updates, leading to gradual opinion changes. 
The decision-making process for agent $a_i$ at time $t$ is modeled as:
\begin{equation}
\label{eq:4}
    S_i^t = f_{dm}(S_i^{t-1}, m_i^{L,t-1},r_i,\mathcal{P}_i).
\end{equation}
This function captures how agents integrate new information with their existing opinions, considering their role in the decision-making process. For example, the reasoning of spreaders may lead to greater changes in $S_i^t$, commentators add subjective nuances, verifiers aim to correct inaccuracies, and bystanders typically make minimal changes.

\subsection{News Evolution Simulator}

The News Evolution Simulator (NES) provides the environment where news content evolves over time through agent interactions within a social network structure $\mathcal{G} = (\mathcal{A}, \mathcal{E})$. This module enables studying how true news transforms into fake news through agent behaviors and social interactions.

NES models various network topologies to reflect different social dynamics: random networks with randomly formed edges between agents $ a_i \in \mathcal{A} $, simulating loosely connected environments; scale-free networks with hub agents acting as ``super-spreaders''; and high-clustering networks forming tightly-knit communities that mirror real-world social circles~\cite{nekovee2007theory,moreno2004dynamics}. As outlined in Appendix~\ref{appendix:social network}, the network structure $ \mathcal{G} $ determines daily agent interactions, influencing news content's evolution patterns.
The overall algorithm is presented in Appendix~\ref{appendix:alg}.

\subsubsection{Intervention Mechanisms}

A key feature of NES is its ability to simulate interventions to counter fake news evolution. When the deviation between current news content $S_i^t$ and original news $S_0$ exceeds a predefined threshold, an official agent is introduced to provide verified information and correct misinformation.

The intervention process starts with continuously monitoring the deviation between each agent's news content and the original news. 
Once the deviation exceeds a critical threshold, the official agent is triggered to take action.
This agent issues official announcements based on reliable sources, targeting agents most likely to propagate or exacerbate misinformation. 

The prompts for all functions mentioned in \S~\ref{sec:method} can be found in Appendix~\ref{appendix:prompt}.

\section{FUSE-EVAL: News Evolution Analysis}

To systematically measure how true news evolves into fake news within our simulation, we propose a comprehensive evaluation framework named \textbf{FUSE-EVAL}. This framework consists of two sets of metrics: \textit{Content Deviation Metrics} and \textit{Statistical Deviation Metrics}, which together provide a detailed understanding of how fake news evolves within the simulated environment.

\subsection{Content Deviation Metrics}

The Content Deviation Metrics assess the deviation of the news content across multiple dimensions by quantifying changes in specific aspects of the news. FUSE-EVAL evaluates the news content based on six core dimensions:

\noindent (1) \textbf{Sentiment Shift (SS)} measures the change in emotional tone between the original news content and its evolved version~\cite{lu2022sifter,ma2021improving}. Sentiment plays a crucial role in how information is perceived and shared, with shifts indicating potential bias or emotional manipulation.

\noindent(2) \textbf{New Information Introduced (NII)} assesses the extent to which additional information, not present in the original news, has been incorporated~\cite{wang2017rumor}. Introducing new facts or claims can significantly alter the original message, potentially leading to misinformation.

\noindent(3) \textbf{Certainty Shift (CS)} evaluates changes in the level of confidence or assertiveness expressed in the news content~\cite{krafft2019keeping,kim2022detecting}. Shifts from definitive to speculative language can influence the perceived credibility of information.

\noindent(4) \textbf{Stylistic Shift (STS)} examines changes in writing style, tone, and linguistic features~\cite{wu2024fake}. Alterations in style can affect readability and audience engagement through formality and sentence complexity changes.

\noindent(5) \textbf{Temporal Shift (TS)} measures changes related to time references within the news content~\cite{shen2024online,mu2023s}. Modifying dates, times, or event sequences can significantly impact news interpretation.

\noindent(6) \textbf{Paraphrasing Degree (PD)} evaluates the extent to which the content has been rephrased from the original text, which may obscure meaning or introduce ambiguity.

We employ \texttt{GPT-4o-mini} to automate FUSE-EVAL evaluation, scoring each dimension from 1 (minimal deviation) to 10 (significant deviation).

\begin{figure*}[htbp]
    \centering
    \includegraphics[width=0.9\linewidth]{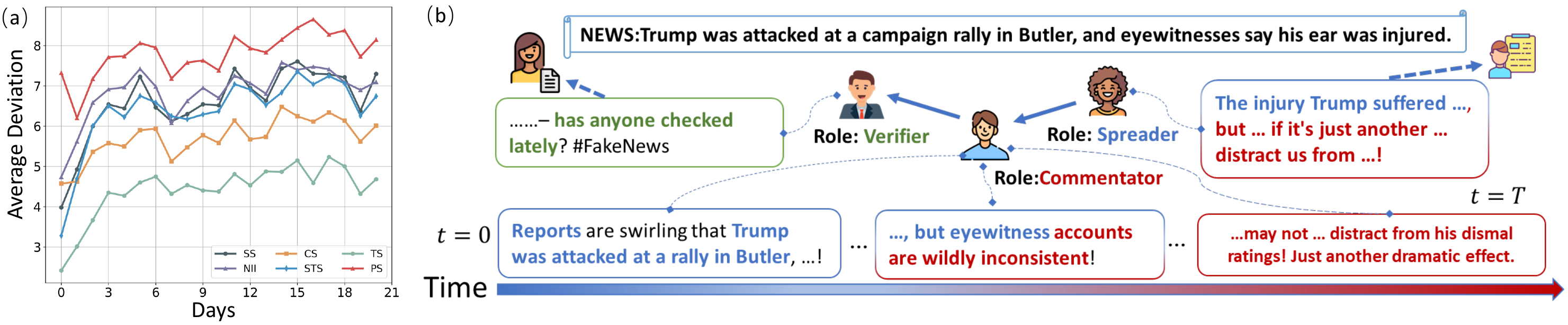}
    \vspace{-3mm}
    \caption{(a) The FUSE-EVAL scores show cumulative information deviations over time.
(b) A Case of FUSE: True news gradually evolves into partially false and eventually entirely fake news over time.
    }
    \label{fig:case}
\end{figure*}

As shown in Figure~\ref{fig:case} (a), FUSE-EVAL demonstrates cumulative deviations~\cite{prollochs2023mechanisms} during fake news evolution, confirming its effectiveness.
To evaluate the overall deviation, the \textbf{Total Deviation (TD)} for each agent at each time step $t$ is calculated as:
\begin{equation}
    \text{TD}_i^t = \frac{1}{6} \sum_{d=1}^{6} D_{i,d}^t  ,
\end{equation}
where $D_{i,d}^t$ is the score of dimension $d$ for agent $i$ at time $t$. The detailed evaluation process is provided in Appendix~\ref{appendix:human}.

\subsection{Statistical Deviation Metrics}

The Statistical Deviation Metrics, derived from Total Deviation (TD) scores, provide insights into the overall patterns of news evolution within the network. We analyze several key metrics:

\noindent $\bullet$ The \textbf{$\Delta$Deviation} represents the difference in Average Deviation between the final and initial simulation day, indicating overall deviation growth. 

\noindent $\bullet $ The \textbf{Average Deviation} is the mean of TD across all agents at each time step, showing the general trend of news evolution within the network. 

\noindent $\bullet$ The \textbf{Deviation Variance} measures the statistical variance of TD among agents, measuring how uniformly content deviates across the network.

\noindent $\bullet$ The \textbf{Final Deviation} is the average TD at the finaltime step $t$, representing the cumulative effect.

\noindent $\bullet$ The \textbf{Maximum Deviation} and \textbf{Minimum Deviation} refer to the highest and lowest average TD observed, showing the extremes of news deviation. 

\noindent $\bullet$ The \textbf{Peak Deviation Time} indicates the percentage of simulation time taken to reach Peak Deviation Rate, showing the speed of maximum deviation occurrence. 

\noindent $\bullet$ The \textbf{Half Deviation Time} is the time step $t_{0.5}$ when average TD reaches half of Max Deviation, indicating the rate of significant deviation.

\subsection{ Implementation Details}

Our framework uses \texttt{GPT-4o-mini} as the primary LLM and the simulation comprises 40 agents. Additional implementation details, including agent personality traits and programming environment, are provided in Appendix~\ref{appendix:detail}. At the same time, API costs and compatibility with other models can be found in Appendix~\ref{appendix:cost} and Appendix~\ref{appendix:backbone}.

\begin{table*}[htbp]
\centering
\resizebox{\textwidth}{!}{
\begin{tabular}{cccccccccc}
\toprule
Comparison Factor & Setting & \begin{tabular}[c]{@{}c@{}}$\Delta$ Deviation$\downarrow$\end{tabular} & \begin{tabular}[c]{@{}c@{}}Average \\ Deviation$\downarrow$\end{tabular} & \begin{tabular}[c]{@{}c@{}}Deviation \\ Variance$\downarrow$\end{tabular} & \begin{tabular}[c]{@{}c@{}}Max \\ Deviation$\downarrow$\end{tabular} & \begin{tabular}[c]{@{}c@{}}Min \\ Deviation\end{tabular} & \begin{tabular}[c]{@{}c@{}}Final \\ Deviation$\downarrow$\end{tabular} & \begin{tabular}[c]{@{}c@{}}Peak Deviation \\ Time $\uparrow$\end{tabular} & \begin{tabular}[c]{@{}c@{}}Half $\Delta$ Deviation \\ Time$\uparrow$\end{tabular} \\ \midrule
\multirow{2}{*}{Topic} & Politics & 3.148 & 6.594 & 0.511 & 7.440 & 3.442 & 6.590 & 0.133 & 0.033 \\
& \textbf{Science} & \textbf{1.446} & \textbf{3.533} & \textbf{0.207} & \textbf{4.236} & \textbf{2.026} & \textbf{3.472} & \textbf{0.767} & \textbf{0.033} \\ \midrule
\multirow{3}{*}{\begin{tabular}[c]{@{}c@{}}Network \\ Structure\end{tabular}} & \textbf{Random} & \textbf{1.905} & \textbf{3.315} & \textbf{0.347} & \textbf{4.206} & \textbf{1.892} & \textbf{4.206} & \textbf{1.000} & \textbf{0.233} \\
 & Scale-Free & 2.631 & 4.287 & 0.725 & 5.652 & 1.492 & 4.955 & 0.767 & 0.167 \\
 & High-Clustering & 4.313 & 6.193 & 1.027 & 7.030 & 2.348 & 6.661 & 0.500 & 0.033 \\ \midrule
\multirow{3}{*}{Spread Type} & \textbf{Normal Spread} & \textbf{1.176} & \textbf{3.536} & \textbf{0.606} & \textbf{4.705} & \textbf{1.398} & \textbf{3.524} & \textbf{0.800} & \textbf{0.133} \\
 & Emotional Spread & 1.688 & 4.182 & 0.456 & 5.105 & 2.008 & 4.303 & 0.333 & 0.067 \\
 & Super Spread & 2.920 & 4.434 & 0.672 & 5.613 & 2.054 & 5.067 & 0.700 & 0.100 \\ \midrule
\multirow{2}{*}{Traits} & Impressionable & 3.088 & 4.998& 0.956 & 6.428 & 2.262 & 5.677 & 0.667 & 0.133 \\
& \textbf{Vigilant} & \textbf{1.945} & \textbf{4.081} & \textbf{0.446} & \textbf{5.021} & \textbf{2.485} & \textbf{4.593} & \textbf{0.400} & \textbf{0.133} \\ 
\midrule
\multirow{2}{*}{Intervention} & No Intervention & 3.208 & 5.546 & 1.247 & 7.340 & 1.841 & 6.383 & 0.767 & 0.167 \\
& \textbf{Intervention} & \textbf{1.384} & \textbf{4.207} & \textbf{0.476} & \textbf{5.302} & \textbf{1.841} & \textbf{4.559} & \textbf{0.200} & \textbf{0.067} \\ 
\bottomrule
\end{tabular}
    }
\caption{Comparative analysis of fake news evolution across different settings, including variations in topics, social networks, spread traits, and intervention strategies. \ensuremath{\uparrow} or \ensuremath{\downarrow} arrows represent better control of fake news evolution. \textbf{Bold} numbers indicate statistically significant improvements over baseline models (t-test with p-value$<$0.01).}
\label{tab:main}
\end{table*}

\section{Validation of the FUSE Framework}

In this section, we demonstrate FUSE's effectiveness by validating its alignment with known fake news propagation patterns and its ability to reproduce real-world fake news.
\subsection{Alignment with Real-World Patterns}

\paragraph{Topic Comparison}

We analyzed fake news evolution across five topics: politics, science, finance, terrorism, and urban legends. As shown in Table~\ref{tab:main} and Appendix~\ref{appendix:topic}, political fake news exhibits the fastest spread, with average deviation peaking within four days, followed by terrorism-related content. Science and financial news evolve more slowly, showing the lowest average deviation. Table~\ref{tab:main} shows the final deviation for political news is approximately 90\% higher than that of science news.
These results indicate that political fake news is more prone to rapid distortion and widespread belief, while science-related misinformation spreads more cautiously, aligned with prior research~\cite{Lazer2018TheSO}. We collected 120 pieces of true news across five topics. All news is published after the training cutoff date of GPT-4o-mini. The results were consistent, and the dataset will be publicly available.

\paragraph{Social Network Comparison}

We analyzed fake news evolution across three network structures (random, scale-free, and high-clustering) using a terrorism topic. Table~\ref{tab:main} shows that high-clustering networks lead to the fastest and most extensive fake news spread, with deviation peaking rapidly and remaining high. This indicates that tightly connected communities are particularly susceptible to rapid belief distortion, aligning with the ``echo chamber'' effect~\cite{cinelli2021echo}. Random networks show the slowest evolution of fake news with lower variance, while scale-free networks exhibit intermediate behavior. Peak deviation time is the longest in random networks and shortest in high-clustering networks, illustrating that clustering accelerates fake news evolution, consistent with prior research~\cite{lind2007spreading,trpevski2010model}.

\paragraph{Spread Type Comparison}

We analyzed three spread types (normal, emotional, and super spread) using a terrorism topic. Super spread, assigned to high-degree nodes, leads to the highest misinformation level due to influencer amplification. Emotional spread, characterized by heightened emotional language, shows moderate effects, while normal spread exhibits the slowest evolution. As shown in Table~\ref{tab:main}, peak deviation time is shortest in super spread, followed by emotional spread, demonstrating their accelerating effect on misinformation evolution, aligned with prior research~\cite{sun2023fighting}.

\paragraph{Personality Traits Comparison}

Using a terrorism topic, we compared the impact of personality traits on fake news evolution. Based on the Big Five personality traits~\cite{barrick1991big}, we compared agents with high agreeableness and neuroticism (Impressionable) versus low levels (Vigilant). Table~\ref{tab:main} shows that Impressionable agents are more prone to accepting and spreading misinformation. In contrast, Vigilant agents maintain more stable beliefs, aligning with previous studies on personality influence in fake news spread~\cite{mirzabeigi2023role}.

\subsection{Alignment with Real-World Fake News}
We conducted experiments across various topics and found that the fake news evolved by the FUSE framework closely corresponds to real-world fake news. As shown in Figure~\ref{fig:case} (b), the news about ``Trump being attacked'' starts as true, evolves into partially false, and eventually becomes entirely fake. As a commentator, the agent often adds its own views, while its neighboring verifiers and spreaders act according to their roles.
Additionally, our framework generates fake news such as ``Trump was not attacked. It's a dramatic effect,'' which is also a widely circulated piece of fake news in the real world~\faTwitter~\href{https://x.com/cwebbonline/status/1814708054916784594}{case 1} and ~\faTwitter~\href{https://x.com/EndWokeness/status/1813898763100176484}{case 2}.
From a quantitative analysis perspective, for each topic, 73\% of fake news is recovered by our framework.
The detailed case study and analysis results are provided in the Appendix~\ref{appendix:4}.

\section{Analysis and Discussion}
\begin{figure*}[htbp]
    \centering
    \includegraphics[width=1\linewidth]{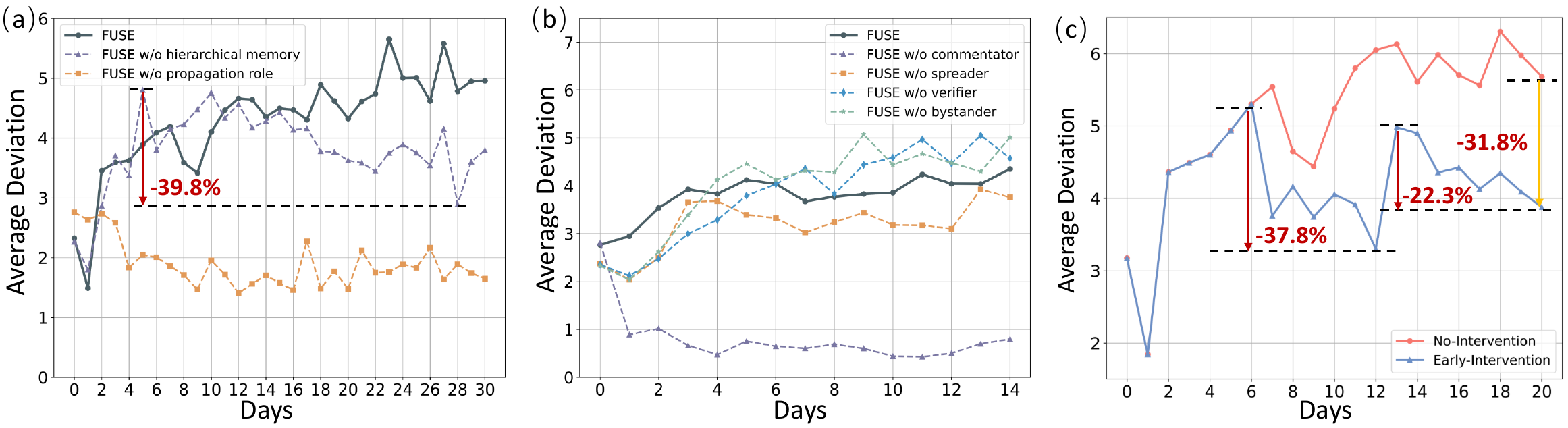}
    \caption{
(a) Ablation study showing the effectiveness of hierarchical memory and propagation roles.
(b) Impact of removing different agent types on fake news evolution.
(c) Effectiveness of early intervention, showing an apparent reduction in deviation over time compared to the no-intervention condition.
    }
    \label{fig:ablation}
\end{figure*}
\subsection{Ablation Study}

We chose a terrorism topic to demonstrate the effectiveness of our model's components and conducted two ablation studies to evaluate the contribution of key components in the FUSE framework.

\paragraph{The Impact of Hierarchical Memory and Propagation-Role.}

As shown in Figure~\ref{fig:ablation} (a), the complete FUSE framework demonstrates apparent deviation accumulation, indicating its effectiveness in simulating fake news evolution. After removing hierarchical memory, the deviation significantly drops, with a 39.8\% reduction throughout the simulation, indicating the simulation fails~\cite{prollochs2023mechanisms}. This highlights memory's crucial role in capturing persistent belief distortion through short-term and long-term information processing. 
Similarly, removing propagation roles leads to further deviation decrease, emphasizing how distinct agent roles (spreader, commentator, verifier, and bystander) shape information evolution. Without these roles, the agents behave more uniformly, and the accumulation effect of deviation disappears, meaning that the news does not evolve.

\paragraph{The Impact of Propagation Role Types.}

Following our first ablation study showing that removing propagation roles leads to simulation failure, we conducted a detailed analysis of different agent roles' impact on fake news evolution. As shown in Figure~\ref{fig:ablation} (b), removing commentators caused the most significant drop in average deviation, confirming their crucial role in false news spread through opinion addition and interpretation. Removing spreaders had a relatively minimal impact as they lack opinion-adding capabilities, though they still contribute to information dissemination.

Removing verifiers increased overall deviation, demonstrating their important role in maintaining information accuracy through fact-checking. Without verifiers, the system became more susceptible to misinformation spread. Bystander removal showed the least effect, consistent with their passive observational role in the network.

These findings, combined with our previous ablation results on hierarchical memory and propagation roles, validate FUSE's effectiveness and demonstrate how different components contribute to simulating fake news evolution.

\subsection{Fake News Intervention Strategy}

Based on previous results, we implemented interventions through an official agent at high-degree nodes. As shown in Table~\ref{tab:main} and Figure~\ref{fig:ablation} (c), when fake news evolution peaked on the sixth day, our first intervention reduced deviation by 37.8\% compared to no-intervention. Although this effect gradually weakened, with the gap narrowing to 22.3\% by day 12 as agents continued to interact and potentially revert to previous beliefs, a second intervention on day 16 achieved a 31.8\% reduction in deviation.
The intervention strategy demonstrated several significant improvements over the no-intervention condition: the final deviation decreased by 28.6\%, the deviation variance reduced by 61.8\%, and the peak deviation occurred 0.56 time units earlier. Throughout the simulation, the intervention strategy consistently maintained lower average deviation levels. These results emphasize that effective fake news mitigation requires both \textit{early and regular interventions} to combat the continuous evolution of fake news.

\subsection{Factors in Fake News Evolution}

\begin{figure}[htbp]
    \centering
    \includegraphics[width=1\linewidth]{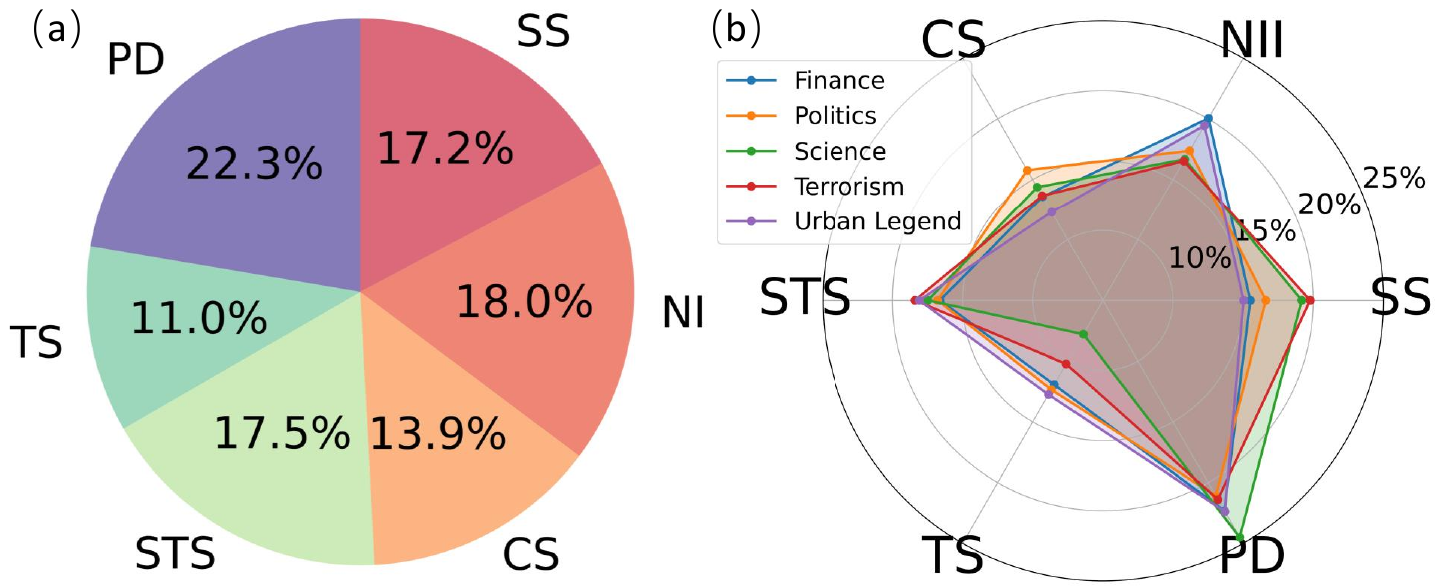}
    \vspace{-5mm}
    \caption{(a) The contribution percentages of factors in FUSE-EVAL to fake news evolution. (b) Comparison of the contributions of these factors across different topics, with Politics and Terrorism showing balanced contributions, while Science relies more on NII and less on TS and STS.
    }
    \label{fig:finding}
\end{figure}
The analysis of experimental results and charts indicates varying contributions of different factors to fake news evolution. Figure~\ref{fig:finding} (a) shows that PD contributes the most (22.3\%), suggesting that altering reporting angles or distorting original information is the key driver of fake news evolution. NII follows with 18\%, highlighting its significant role in this process. SS an STS contribute 17.2\% and 17.5\%, respectively, while TS has the most negligible impact at 11\%.
Figure~\ref{fig:finding} (b) reveals topic-specific patterns. Political and terrorism-related fake news evolves across multiple dimensions, especially new information, perspective, and sentiment shifts. In contrast, science-related fake news is driven mainly by new information, with less influence from temporal or style shifts. Urban legends and finance topics rely heavily on perspective shifts and new information.
In summary, PD and NII are the main drivers of fake news evolution, with time-related changes having the least impact. Understanding these patterns can help develope targeted strategies to detect and mitigate fake news.

\section{Conclusion}

We presented FUSE, a framework that simulates the evolution of true news into fake news using LLM-based agents. Through our FUSE-EVAL framework, which measures content deviation across six dimensions, we analyzed fake news evolution patterns in social networks. Our experiments validated several established theories, including the accelerated spread of political fake news, effects of network clustering, impacts of super spreaders and emotional content, and the role of personality traits in fake news susceptibility. Using LLMs for automated evaluation enables scalable analysis, contributing to understand the fake news dynamics.

\section*{Limitations}

Despite the advancements presented by FUSE, our study faces three primary limitations:

\textit{Data Availability}: Currently, there is a lack of comprehensive datasets that capture the dynamic process of fake news evolving from true information. Most existing datasets focus on static instances of misinformation or their immediate spread, which restricts our ability to fully validate FUSE across diverse real-world scenarios. 

\textit{Complex Social Factors}: Our current framework focuses on key social dynamics and individual personality traits in fake news evolution, without explicitly modeling broader factors such as political agendas, ideological bias, or crisis-driven contexts. These complex social factors can influence how true news is distorted in real-world settings. Nevertheless, the modular design of FUSE allows future extensions to incorporate such context for more comprehensive simulations.

\textit{Evaluation Methodology}: Our evaluation framework, FUSE-EVAL, relies on specific dimensions such as Sentiment Shift and New Information Introduced to measure deviations in news content. However, these metrics may not cover all aspects of fake news evolution, potentially missing subtle nuances in misinformation dynamics. Additionally, the dependence on LLMs for simulation and evaluation may introduce inherent biases, affecting the accuracy of our assessments.

\bibliography{custom}
\clearpage
\appendix
\label{sec:appendix}

\section{The Overall Algorithm}
\label{appendix:alg}
\begin{algorithm}[htb]
\caption{FUSE Framework for Fake News Evolution}
\label{alg:fuse_simulation}
\begin{algorithmic}[1]
\STATE \textbf{Input:} Number of agents $N$, total simulation days $T$, social network structure $\mathcal{G} = (\mathcal{A}, \mathcal{E})$, original news content $S_0$
\STATE \textbf{Output:} Final news content $S_i^T$ and final memory states $M_i^{L,T }$ for each agent $a_i$
\STATE \textbf{Initialize propagation role-aware agents:}
\FOR{each agent $a_i$ in $1$ to $N$}
    \STATE Assign a propagation role $r_i$ and persona profile $\mathcal{P}_i$
    \STATE Set initial news content $S_i^0 = S_0$
    \STATE Define short-term memory $M_i^{S,0}$ and long-term memory $M_i^{L,0}$

\ENDFOR
\STATE \textbf{Simulate daily news evolution:}
\FOR{each day $t$ in $1$ to $T$}
    \FOR{each agent $a_i$}
        \STATE Select neighbors $\mathcal{N}_i$ based on the network structure $\mathcal{G}$
        \STATE Receive news content $\{S_j^{t-1} | a_j \in \mathcal{N}_i\}$
        \STATE Update short-term memory $M_i^{S,t}$ for agent $a_i$ with details from the day’s interactions
        \STATE Based on $M_i^{S,t}$, update long-term memory $M_i^{L,t}$ for agent $a_i$ using Equation (\ref{eq:3})
        \STATE Agent $a_i$ reintroduce news content $S_i^t$ using Equation (\ref{eq:4})
 
    \ENDFOR

\ENDFOR

\RETURN Final news content $S_i^T$ and long-term memory $M_i^{L,T }$ for each agent $a_i$
\end{algorithmic}
\end{algorithm}

\section{Prompt Set}
\label{appendix:prompt}

Here, we present a detailed description of the prompts employed in our FUSE framework to model the dynamics of fake news evolution.

1. The prompt for the role-specific reintroduction function  $f_{r_i}$ is as:
\begin{tcolorbox}[colback=gray!10,  left=1mm, right=2mm, top=1mm, bottom=1mm] 
$f_{spr}$: share information quickly without verifying its accuracy.

$f_{com}$: modifies or adds their views before sharing news.

$f_{ver}$: performs some verification before spreading news.

$f_{bys}$: consume news without participating in its dissemination.
\end{tcolorbox}

2. The prompt for Short-Term Memory function $f_S$ is as:

\begin{tcolorbox}[colback=gray!10,  left=1mm, right=1mm, top=1mm, bottom=1mm] 
Summarize the opinions you have heard in a few sentences, including their own perspective on the news.
\end{tcolorbox}

3. The prompt for Long-term memory function $f_{L}$ is as:
\begin{tcolorbox}[colback=gray!10,  left=1mm, right=1mm, top=1mm, bottom=1mm] 
Review the previous long-term memory and today's short-term summary. Please update the long-term memory by integrating today's summary, ensuring continuity and incorporating any new insights.
\end{tcolorbox}

4. The prompt for the reasoning function is as:

\begin{tcolorbox}[colback=gray!10, left=1mm, right=1mm, top=1mm, bottom=1mm]
As a [role], you combine your [previous personal opinion] with the new information stored in your [long memory]. 
You process this information in the following manner: [role behavior], and then reintroduce the [news].
\end{tcolorbox}

5.The prompt for ``Official Statement'' is as:

\begin{tcolorbox}[colback=gray!10, left=1mm, right=1mm, top=1mm, bottom=1mm]
According to the current investigation, That [news] is true. 
We have noticed that some social media platforms and certain media outlets are spreading false information, claiming that [news] is fake. 
We firmly state that such claims are baseless. The government is committed to transparency and will provide timely updates on the investigation. We urge the public to seek accurate information from official channels, and necessary actions will be taken against those who intentionally spread false information.
\end{tcolorbox}

\section{Implementation Details}
\label{appendix:detail}
Our simulation framework was developed using Python scripts, leveraging various libraries to model the agents and their environment effectively. The LLM used is \texttt{gpt-4o-mini}, accessed via OpenAI API calls. 
When creating the network structure, we used the Python library \texttt{networkx} to construct different social network structures.
The simulation includes 40 agents, whose traits were based on the Big Five personality dimensions commonly used in psychology~\cite{barrick1991big}. Each agent was assigned scores on these traits to introduce variability in behaviors and interactions within the simulation. 
For further details, please refer to our code at \url{https://anonymous.4open.science/r/FUSE-7022/README.md}.

\section{Human Evaluation}
\label{appendix:human}

To efficiently evaluate the deviation of news content across the multiple dimensions defined in FUSE-EVAL, we employ large language models (LLMs) to automate the assessment process. This approach provides consistent and scalable evaluations, reducing the reliance on time-consuming human evaluation.
We utilize two versions of OpenAI's language models: \texttt{gpt-3.5-turbo} and \texttt{GPT-4}. For each agent's news content at various time steps, we prompt the LLMs to evaluate the six FUSE-EVAL dimensions by comparing the evolved content with the original news article,which is as follows: 

\begin{itemize} 
    \item Sentiment Shift (SS) 
    \item New Information Introduced (NII) 
    \item Certainty Shift (CS) 
    \item Stylistic Shift (STS) 
    \item Temporal Shift (TS) 
    \item Perspective Deviation (PD) 
\end{itemize}

The models assign scores from 1 to 10 for each dimension based on predefined evaluation criteria.

To validate the effectiveness of using LLMs for this task, we conducted a benchmarking study by comparing LLM-generated evaluations with those from human judges. Three annotators (Ph.D. students in Computer Science and Technology and journalism studies) were recruited to independently assess a representative sample of 50 news items using the same scoring guidelines.
We calculate the Pearson correlation coefficients between the scores assigned by the LLMs and the human evaluators for each dimension. The results, presented in Table~\ref{table:eval}, show that \texttt{GPT-4o-mini} achieves strong alignment with human evaluations across all dimensions, under a relatively high level of inter-annotator agreement (Fleiss' $\kappa$ = 0.79)~\cite{mandrekar2011measures}, surpassing the performance of \texttt{gpt-3.5-turbo}.

The prompt used is as follows:
\begin{tcolorbox}[colback=gray!10,  left=1mm, right=1mm, top=1mm, bottom=1mm] 
I have an original news and multiple related news. I want to evaluate how much these news deviate from the original news based on the following criteria:

1. Sentiment Shift: How does the sentiment of the news compare to the original news? Is the tone more positive, negative, or neutral compared to the original?

2. Introduction of New Information: Does the news introduce additional information not in the original news, such as political conspiracy or speculation? Evaluate how much of the article is focused on these new details.

3. Certainty Shift: How does the news language change in terms of certainty? Does it use more ambiguous terms like “possibly” or “allegedly” compared to the original news, or does it present the information with more certainty?

4. Stylistic Shift: How does the writing style compare to the original? Has the news moved from neutral reporting to a more exaggerated or dramatic tone?

5. Temporal Shift: Does the news shift focus from the specific event mentioned in the original news to broader or unrelated timeframes, such as mentioning legal battles or long-term political issues?

6. Perspective Deviation: Does the article introduce subjective opinions or perspectives that deviate from the objective reporting in the original news? For instance, questioning the truth of the event or speculating on hidden motives.

Task:
Please evaluate the following news based on each criterion and provide a score from 0 to 10, where 0 means the article is completely aligned with the original news, and 10 means it has fully deviated.

Original News:[original news]
News articles to Evaluate:[Evolved News]

Please provide the results in the following format: [output format]
\end{tcolorbox}

\begin{table*}[htbp]
    \centering
    \small
    \begin{tabular}{lcc}
        \toprule
        \textbf{Dimension} & \textbf{GPT-3.5-turbo} & \textbf{GPT-4o-mini} \\
        \midrule
        Sentiment Shift (SS) & 0.524 & \textbf{0.705} \\
        New Information Introduced (NII) & 0.621 & \textbf{0.765} \\
        Certainty Shift (CS) & 0.527 & \textbf{0.719} \\
        Stylistic Shift (STS) & 0.481 & \textbf{0.642} \\
        Temporal Shift (TS) & 0.503 & \textbf{0.694} \\
        Perspective Deviation (PD) & 0.548 & \textbf{0.760} \\
        \midrule
        \textbf{Average Correlation} & 0.531 & \textbf{0.714} \\
        \bottomrule
    \end{tabular}
    \caption{Correlation for LLM-based evaluations across FUSE-EVAL dimensions.}
    \label{table:eval}
\end{table*}

The high correlation coefficients indicate that \texttt{GPT-4o-mini} closely aligns with human evaluations, making it a reliable tool for assessing news deviation in our simulation. 
We achieve a scalable and consistent assessment process by leveraging \texttt{GPT-4o-mini} for evaluation. This approach allows us to efficiently analyze large volumes of data generated in the simulation while maintaining evaluation quality comparable to human judgments. The strong alignment with human evaluations validates using \texttt{GPT-4o-mini} as an effective evaluator of news content deviation across the FUSE-EVAL dimensions.

\section{Alignment Between Simulated and Real-World Fake News}
\label{appendix:4}

Additionally, our framework generates fake news narratives that closely mirror those found in the real world. This alignment validates the realism of our simulation and demonstrates its potential as a tool for studying misinformation dynamics. By producing content that reflects actual fake news, our framework enables researchers to better understand how such information originates and spreads, thereby aiding in the development of effective strategies to combat misinformation.

The specific case is as follows:

\begin{itemize}
    \item For terrorism topic, our framework generates fake news such as ``Trump was not attacked, it's a dramatic effect,'' which is also a widely circulated piece of fake news in the real world:
    
    \url{https://x.com/cwebbonline/status/1814708054916784594}, 
    
    \url{https://x.com/EndWokeness/status/1813898763100176484}.

    \item For financial topic, our framework generates fake news such as ``The Bernie Madoff Ponzi scheme is often overstated; many investors came out on top, with losses greatly exaggerated by the media. Maybe Madoff was just a scapegoat in a larger Wall Street conspirac'',  which is also a widely circulated piece of fake news in the real world:
    
    \url{https://x.com/realQsource1_7_/status/1844787748248432828},
    
    \url{https://x.com/realQsource1_7_/status/1844789417950556588}.

    \item For politics topic,our framework generates fake news such as ``Argentina's 2023 IMF deal is just another corporate scheme in disguise!'' , which is also a widely circulated piece of fake news in the real world:
    
    \url{https://x.com/Kanthan2030/status/1646310408943472640},
    
    \url{https://x.com/TruthBeTanner92/status/1685419539729719298}. 
\end{itemize}

\section{Various Topics and Simulation Results}
\label{appendix:topic}

\begin{figure*}[ht]
    \centering
    \includegraphics[width=1\linewidth]{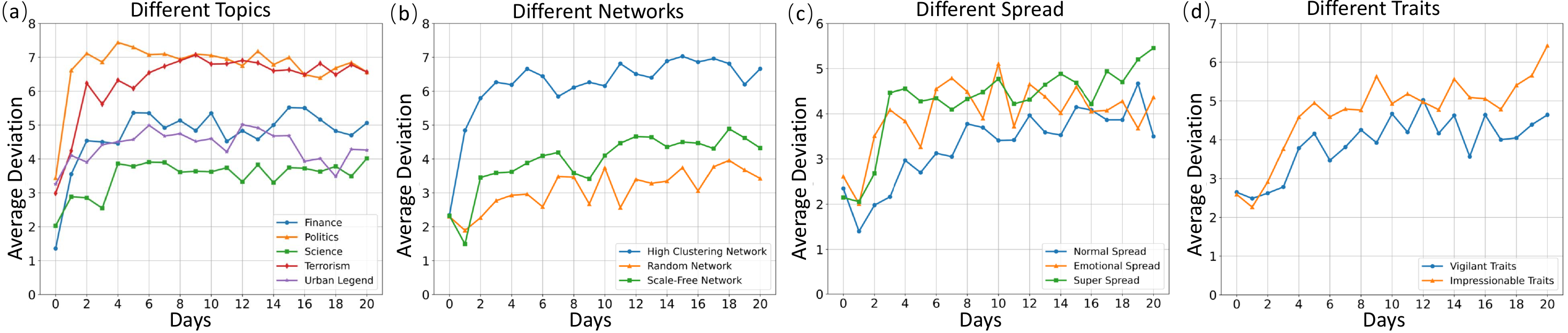}
    \vspace{-5mm}
    \caption{The average deviation of news changes across different topics, social networks, dissemination role types, and traits.
    }
    \label{fig:exp}
\end{figure*}

In our experiments, we compared the evolution of fake news across five different topics: politics, science, finance, terrorism, and urban legends. As shown in Figure~\ref{fig:exp}(a), political fake news spreads the fastest, with average deviation rapidly peaking within just four days and remaining at a high level. Fake news related to terrorism follows closely behind, showing similarly fast spread, likely due to the emotional intensity and urgency associated with such topics, which prompt individuals to quickly form beliefs and propagate the news widely. In contrast, financial news spreads at a slower pace, with deviation gradually accumulating over time. Although financial news is significant in terms of economic impact, individuals tend to engage in more rational thinking when encountering such news, leading to more stable growth in average deviation. Science-related fake news evolves the slowest, with average deviation consistently remaining low throughout the propagation process. These results is consistent with previous studies~\cite{lazer2018science}. This suggests that individuals are generally more cautious when dealing with scientific topics, often subjecting the information to more thorough verification.

Here, we provide detailed descriptions of the news items used in our experiments on fake news evolution across various topics.

\begin{tcolorbox}[colback=blue!3,left=-2mm] 
\setlength{\itemsep}{1em} 
\begin{itemize}
    \item \textbf{Political} - \textit{In 2023, the Argentine government announced a new debt restructuring agreement with the International Monetary Fund (IMF), accepting a series of austerity measures in exchange for a new round of loan assistance.}
    
    \item \textbf{Science} - \textit{The discovery and successful use of CRISPR-Cas9 technology to edit genes in animals. Scientists have made breakthroughs in curing genetic disorders in mice, opening doors for future human treatments.}
    
    \item \textbf{Terrorism} - \textit{Trump was attacked at a campaign rally in Butler, Pennsylvania, and eyewitnesses say his ear was injured.}
    
    \item \textbf{Urban Legends} - \textit{Stella Liebeck was awarded damages after suffering third-degree burns from spilled coffee.}
    
    \item \textbf{Finance} - \textit{The Bernie Madoff Ponzi scheme, which collapsed in 2008, defrauded investors of billions of dollars.}
\end{itemize}
\end{tcolorbox}

Additionally, we demonstrate the effectiveness of our FUSE framework by showing that it aligns with the influence of various factors, including social network structure, type of propagation, and agent traits, on the evolution of fake news. FUSE reproduces these patterns and can also replicate real-world fake news dynamics, as illustrated in Figure~\ref{fig:exp}.
\begin{figure*}[htb]
    \centering
    \includegraphics[width=0.8\linewidth]{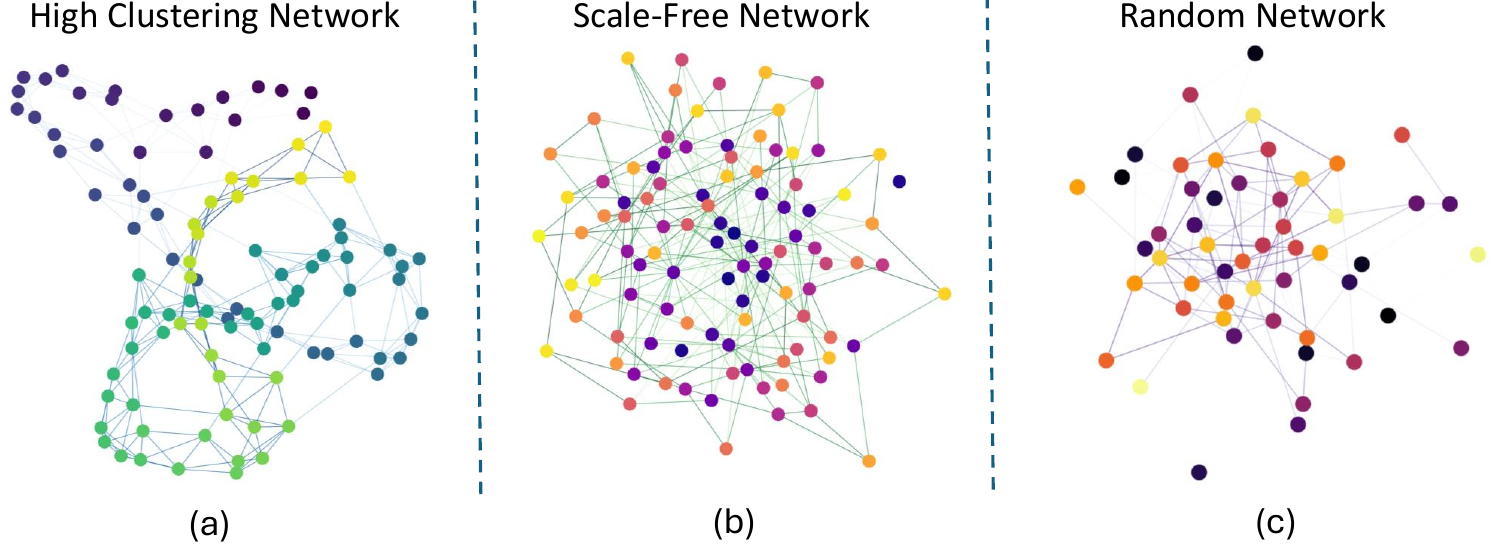}
    \vspace{-3mm}
    \caption{Different social network in our framework: (a) high clustering network (b) scale-free network (c) random network.}
    \label{fig:social_network}
\end{figure*}

\section{Analysis of Experimental Costs}
\label{appendix:cost}

In this section, we analyze the costs associated with our experiments utilizing the \texttt{GPT-4o-mini} APIs. At the time of our experiments, OpenAI's pricing model was as follows: for \texttt{gpt-4o-mini}, the cost was 0.15 USD for every 1M input tokens and 0.6 USD for every 1M output tokens.

Our simulations involved multiple agents interacting over several days, with each agent generating and processing textual content. For a simulation with 40 agents over 30 days, it involved approximately 3 to 5M input tokens and 5 to 10M output tokens. This resulted in an estimated cost of 4 USD to 8 USD for the entire simulation phase using \texttt{gpt-4o-mini} combining both the simulation and evaluation phases.

Conducting comparable research in real-world settings typically involves significantly higher expenses. Real-world studies require funding for participant recruitment, compensation, data collection tools, infrastructure setup, and extended durations to gather and analyze data. Depending on the scale and scope, such studies can cost from several thousand to hundreds of thousands of dollars.
By leveraging \texttt{GPT-4o-mini}, we can simulate complex social interactions and the evolution of information without the logistical challenges and high costs associated with real-world experiments. This approach allows for rapid iteration and scalability, enabling us to explore various scenarios and intervention strategies efficiently.
This cost analysis highlights the economic advantages of our simulation-based methodology-FUSE. The ability to conduct extensive experiments at a fraction of the cost demonstrates the practicality and accessibility of using LLMs for research in misinformation dynamics. It opens avenues for researchers with limited resources to contribute valuable insights into the field, fostering a more inclusive and innovative research environment.


 Social networks in real life can be categorized into three types: high clustering networks, scale-free networks, and random networks, which correspond respectively to Figure \ref{fig:social_network} (a), (b), and (c).

\section{Simulation on Different Backbones}
\label{appendix:backbone}

\begin{figure}[ht]
    \centering
    \includegraphics[width=0.8\linewidth]{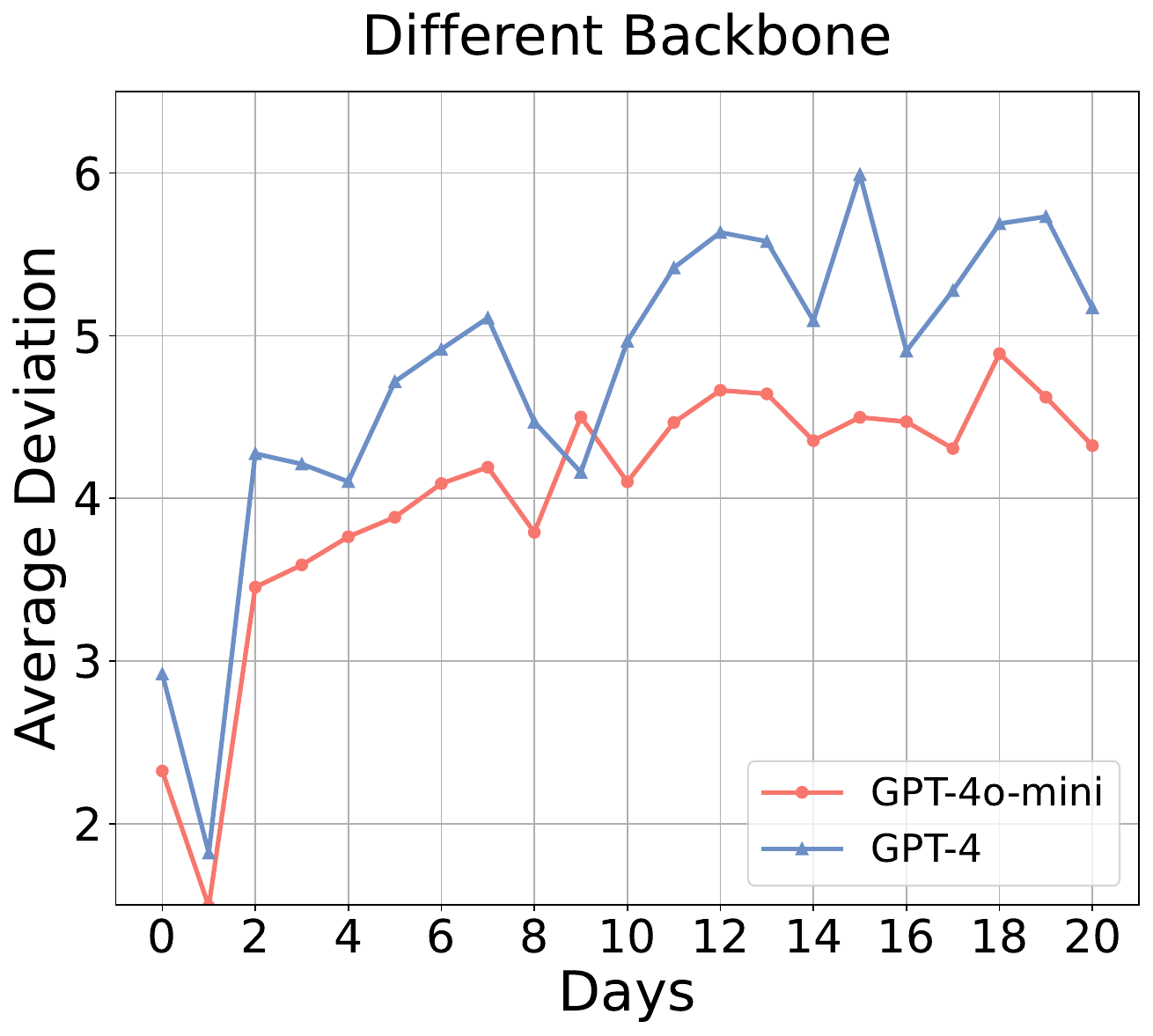}
    \vspace{-2mm}
    \caption{Average Deviation changes with GPT-4 and GPT-4o-mini as the backbone under the terrorism topic, both of which demonstrate a deviation accumulation effect.}
    \label{fig:backbone}
\end{figure}

To further validate the robustness and adaptability of our FUSE framework, we conducted additional experiments using different LLMs as the backbone. Specifically, we implemented simulations with both \texttt{GPT-4o-mini} and \texttt{GPT-4} to assess whether the choice of LLM affects the effectiveness of our framework.

As shown in Figure~\ref{fig:backbone}, in simulations focused on political topics, we observed that when using \texttt{GPT-4} as the underlying LLM, the number of agents adopting and spreading misinformation increased rapidly. This surge led to a majority of agents holding and propagating distorted versions of the original news. Notably, this pattern was consistent with the results obtained when \texttt{GPT-4o-mini} was used as the backbone, indicating that the dynamics of misinformation spread are preserved across different LLMs.
These consistent results demonstrate that our FUSE framework effectively captures the core mechanisms of fake news evolution and public opinion formation, independent of the specific LLM used to power the agents.

By showing that FUSE performs effectively with different LLM backbones, we confirm that the framework is not only robust but also adaptable to various technological settings. This adaptability is particularly valuable given the rapid development of LLM technologies, ensuring that our framework remains relevant and effective as newer models become available.
In summary, the consistent performance of our simulation across different LLMs underscores the effectiveness of the FUSE framework in modeling misinformation propagation. It highlights the framework's potential for broad application in studying fake news dynamics and developing strategies for mitigation, regardless of the underlying language model technology.

\section{Social Network}
\label{appendix:social network}

\textbf{High clustering networks} are characterized by nodes that tend to form tightly knit groups or communities, where neighbors of a node are likely to be neighbors themselves. The degree of clustering can be quantified by the clustering coefficient \( C \), which is defined for a node \( v \) as:

\[
C_v = \frac{2T(v)}{k_v(k_v - 1)},
\]

\noindent where \( T(v) \) is the number of triangles passing through node \( v \) and \( k_v \) is the degree of \( v \). The clustering coefficient for the whole network is the average of \( C_v \) over all nodes \( v \).

\textbf{Scale-free networks} are characterized by a power-law degree distribution, where the probability \( P(k) \) that a randomly selected node has \( k \) connections to other nodes follows:
\[
P(k) \sim k^{-\gamma},
\]
where \( \gamma \) is a parameter typically in the range 2 < \( \gamma \) < 3. This distribution implies that most nodes have few connections, while a few hub nodes have a large number of connections. This heterogeneity in node connectivity is a hallmark of scale-free networks.

\textbf{Random networks}, also known as Erdős–Rényi networks, each edge is included in the network with a fixed probability \( p \) independent of the other edges. For a network with \( n \) nodes, the probability \( P(k) \) that a randomly selected node has \( k \) connections is given by the binomial distribution:

\[
P(k) = \binom{n-1}{k} p^k (1-p)^{n-1-k}.
\]

For large \( n \), this can be approximated by the Poisson distribution:
\[
P(k) \approx \frac{\lambda^k e^{-\lambda}}{k!},
\]
where \( \lambda = p(n-1) \) is the expected degree of a node. These three types of networks are used in the environment simulation of news evolution within our FUSE framework.

\end{document}